\documentclass[twocolumn,aps,prd,superscriptaddress]{revtex4-1}
\usepackage{amsmath}
\usepackage{bm}
\usepackage{graphicx}
\usepackage{mathrsfs}
\usepackage{multirow}
\usepackage[normalem]{ulem}
\usepackage{graphicx}
\usepackage{booktabs, ctable}
\usepackage{xcolor, color, framed}
\usepackage[colorlinks, linkcolor=red,anchorcolor=green,citecolor=blue]{hyperref}

\begin{document}

\title{Probe charmonium-nucleon interactions in high energy proton-proton collisions}

 \author{Jiaxing Zhao}
 \affiliation{Helmholtz Research Academy Hesse for FAIR (HFHF), GSI Helmholtz Center for Heavy Ion Physics, Campus Frankfurt, 60438 Frankfurt, Germany}
 \affiliation{Institut f\"ur Theoretische Physik, Johann Wolfgang Goethe-Universit\"at,Max-von-Laue-Straße 1, D-60438 Frankfurt am Main, Germany}

\author{Taesoo Song}
\affiliation{GSI Helmholtzzentrum f\"{u}r Schwerionenforschung GmbH, Planckstrasse 1, 64291 Darmstadt, Germany}

\author{Joerg Aichelin}
\affiliation{SUBATECH UMR 6457 (IMT Atlantique,  Universit\'{e} de Nantes, IN2P3/CNRS), 4 Rue Alfred Kastler, F-44307 Nantes, France}
\affiliation{Frankfurt Institute for Advanced Studies, Ruth-Moufang-Strasse 1, 60438 Frankfurt am Main, Germany}

\author{Elena Bratkovskaya}
\affiliation{GSI Helmholtzzentrum f\"{u}r Schwerionenforschung GmbH, Planckstrasse 1, 64291 Darmstadt, Germany}
\affiliation{Helmholtz Research Academy Hesse for FAIR (HFHF), GSI Helmholtz Center for Heavy Ion Physics, Campus Frankfurt, 60438 Frankfurt, Germany}
\affiliation{Institut f\"ur Theoretische Physik, Johann Wolfgang Goethe-Universit\"at,Max-von-Laue-Straße 1, D-60438 Frankfurt am Main, Germany}

\author{Pol Bernard Gossiaux}
\affiliation{SUBATECH, Nantes University, IMT Atlantique, IN2P3/CNRS, 4 rue Alfred Kastler, 44307 Nantes cedex 3, France}

\author{Klaus Werner}
\affiliation{SUBATECH UMR 6457 (IMT Atlantique,  Universit\'{e} de Nantes, IN2P3/CNRS), 4 Rue Alfred Kastler, F-44307 Nantes, France}

\date{\today}

\begin{abstract}
We investigate charmonium production and the charmonium–nucleon correlation function in pp collisions using the EPOS4+CATS framework. For the first time, the emission source of charmonium–proton pairs is dynamically generated and found to be non-Gaussian. This enables a femtoscopic extraction of the charmonium–proton interaction directly from experimental correlation functions. Both ground and excited charmonium states are included. We find that feed-down from excited charmonium states can induce sizable modifications to the observed prompt $J/\psi$-proton correlation function, even when the correlation deviates from unity by less than one, reflecting the stronger interactions of the excited states. 
\end{abstract}

\maketitle

The interaction between charmonia and nucleons (N), particularly in the $J/\psi-N$ system, has been a subject of longstanding interest in both nuclear and hadronic physics. Its importance can be summarized as follows. First, due to its compact size, charmonium serves as an excellent probe of gluon fields inside hadrons and nuclei~\cite{Krein:2020yor}. Unlike light hadrons, charmonium consists exclusively of heavy charm quarks, and its coupling to nucleons proceeds predominantly through multi-gluon exchanges. This provides a unique window into the gluonic structure of the nucleon and into non-perturbative aspects of QCD, including the contribution of the trace anomaly to the nucleon mass~\cite{Kharzeev:1995ij,Ji:1994av}. Second, the charmonium-nucleon interaction is closely related to the structure and properties of hidden-charm pentaquarks, such as the $P_c$ states~\cite{LHCb:2015yax,LHCb:2019kea}. Finally, it is essential for understanding the in-medium modifications of charmonium~\cite{Sibirtsev:2005ex} and for improving their sensitivity as probes of the deconfined QCD medium—the quark–gluon plasma (QGP).

The charmonium–nucleon interaction has been extensively studied with a variety of theoretical approaches, and most results consistently suggest that the interaction is attractive but relatively weak. Within effective field theory approaches, the interaction potential has been investigated using the QCD multipole expansion (QCDME), in which the leading contribution originates from the chromoelectric dipole coupling between the compact charmonium and the gluon fields inside the nucleon~\cite{Peskin:1979va,Bhanot:1979vb,Eides:2017xnt,Voloshin:2007dx,Yeo:2025ufo}. In this picture, the interaction is short-range and attractive, analogous to a van der Waals force in QED, and can be parameterized by the quarkonium chromoelectric polarizability. The resulting potential yields a small but finite scattering length, which is crucial for estimating possible charmonium–nucleon bound states and in-medium modifications of charmonia~\cite{Luke:1992tm,Kharzeev:1994pz}. This framework has been extended to include coupled-channel effects, in which  $J/\psi$–nucleon scattering into open-charm meson–baryon channels and inverse modifies the effective interaction~\cite{Wu:2024xwy}.
Beyond multipole expansion, a dedicated quarkonium–nucleon effective field theory (QNEFT) has been constructed~\cite{TarrusCastella:2018php}, incorporating $S$-wave quarkonium fields, nucleon doublets, and pion triplets in accordance with chiral symmetry, heavy-quark spin symmetry, and CPT invariance. Furthermore, lattice QCD calculations-both in the quenched approximation~\cite{Yokokawa:2006td,Kawanai:2010ru,Kawanai:2010ev} and with unphysically heavy pion masses~\cite{Sugiura:2019pye}—have provided valuable information such as $S$-wave scattering lengths $a_0$ and effective ranges $r_{\rm eff}$. More recently, the HAL QCD Collaboration has performed (2+1)-flavor simulations with nearly physical pion masses, offering new insights into both the $J/\psi-N$ and $\eta_c-N$ interactions~\cite{Lyu:2024ttm}.

Experimentally, the strength of the charmonium–nucleon interaction remains only weakly constrained. Analyses of $J/\psi$ photoproduction off nuclei indicate a small absorption cross section, consistent with a weak interaction and a small scattering length~\cite{GlueX:2019mkq}. Developing effective strategies to probe the charmonium–nucleon interaction in experiments is therefore highly desirable. One promising approach is femtoscopy, which provides access to hadronic interactions as well as information about the space–time structure of the hadron emission source~\cite{Lisa:2005dd,NA49:2007fqa,Li:2008qm,Wiedemann:1996ig,Kisiel:2006is,Xu:2024dnd,Si:2025eou}. The technique connects the measured two-particle correlation to the underlying two-body interaction~\cite{Lisa:2005dd}. This gives us an opportunity to probe the charmonium–nucleon potential via their correlations. 
Many correlations, measured at RHIC and LHC, have demonstrated the feasibility of this method, leading to new insights into nucleon-hyperon and hyperon-hyperon interactions~\cite{STAR:2005rpl,ALICE:2018ysd,ALICE:2019buq,ALICE:2019hdt,STAR:2018uho,STAR:2014dcy,ALICE:2022uso}. Recently it has been  extended to heavy-light system, such as $D-\pi$, $D-K$, and $D-p$ system~\cite{ALICE:2022enj,ALICE:2024bhk,Gwizdziel:2024xpl}. This extension offers a promising opportunity to investigate the charmonium–nucleon system in the upcoming LHC Run 3 and future experiments.

Within the femtoscopy formalism, the experimentally observed correlation function can be expressed theoretically as a convolution of the source function $S(\mathbf{r})$ with the relative two-body scattering wave function $\psi_k(\mathbf{r})$~\cite{Koonin:1977fh},
\begin{eqnarray}
C(k)=\int S({\bm r})|\psi_k({\bm r})|^2d{\bm r},
\label{eq.correlation}
\end{eqnarray}
where $k=|{\bf p}_1^*-{\bf p}_2^*|/2$ is the relative momentum in the center-of-mass frame of the pair, and ${\bf r}$ is the relative distance between the two particles. $\psi_k({\bf r})$ is the two-body scattering wave function.
The charmonium–nucleon correlations via femtoscopy have been explored in several pioneering studies~\cite{Krein:2020yor,Krein:2023azg,Liu:2025oar}, based on the Lednick\'y-Lyuboshits model~\cite{Lednicky:1981su} or phase shifts approach~\cite{Liu:2025oar}. In both cases, the source function is typically approximated by a Gaussian distribution.
For the $J/\psi-N$ system, a more accurate treatment can be achieved by directly solving the Schr\"odinger equation with realistic interaction potentials. Equally important is a realistic description of the emission source. The correlation function is determined by the interplay between the source distribution and the final-state interaction, implying that any extraction of the interaction from measured correlations is inherently source dependent. As recently emphasized in Ref.~\cite{Epelbaum:2025aan}, a consistent treatment of both the source and the interaction is therefore essential for obtaining reliable constraints from femtoscopic observables.

While a Gaussian source often provides a reasonable approximation for light hadrons emitted from a thermalized medium, its applicability to rare probes such as the $J/\psi$ is less evident.
In this paper, we employ for the first time the EPOS4 framework to determine the charmonium--nucleon emission source, which already provides a good description of both light, open heavy-flavor~\cite{Zhao:2024ecc,Zhao:2023ucp}, and quarkonium production~\cite{Zhao:2025cnp}. This allows us to obtain a realistic charmonium–nucleon emission source, thereby significantly improving the precision of femtoscopic constraints on the interaction.

\begin{figure}[!htb]
    \centering
    \includegraphics[width=0.46\textwidth]{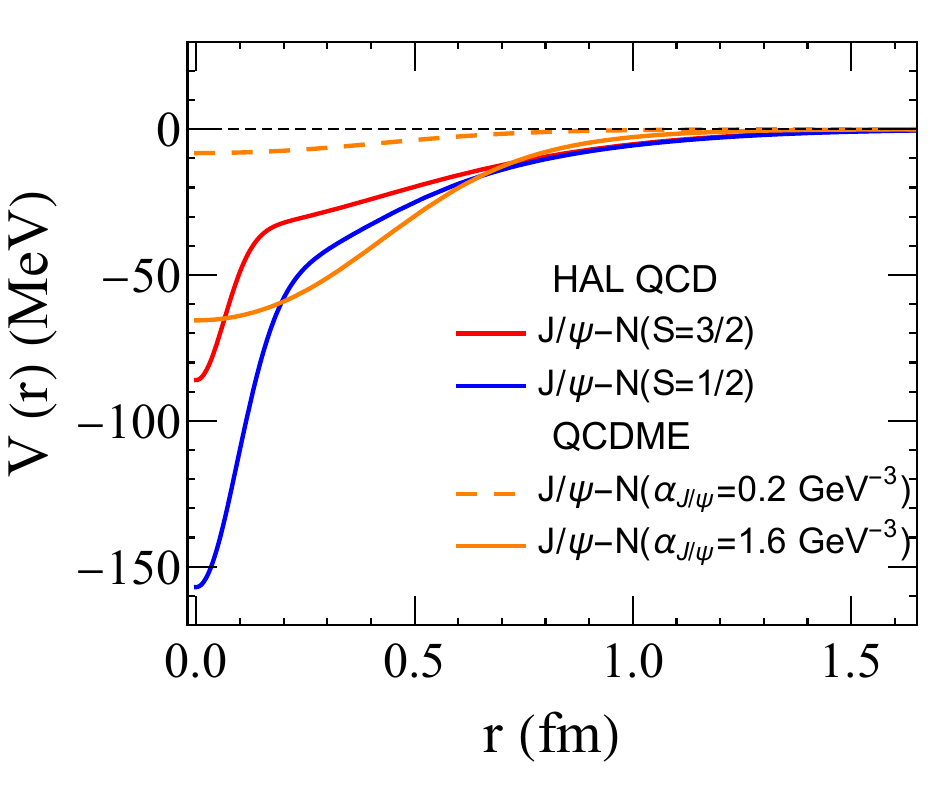}    
    \caption{The interaction potential of $J/\psi-N$ from HAL QCD with spin-3/2 (red) and 1/2 (blue) state and $J/\psi-N$ from the QCDME with perturbative chromoelectric polarizability $\alpha_{J/\psi}=0.2~\rm GeV^{-3}$ (orange-dashed) and non-perturbative one $\alpha_{J/\psi}=1.6~\rm GeV^{-3}$ (orange-solid).}
    \label{fig.pot}
\end{figure}

We first adopt the charmonium–nucleon interaction derived from the QCDME~\cite{Voloshin:2007dx},
\begin{eqnarray}
V(r)=-{1\over 2}\alpha_{\psi}g^2\left({8\pi^2\over bg_s^2}T_\mu^\mu+T^G_{00}\right),
\label{eq.potME}
\end{eqnarray}
where $g$ and $g_s$ denote the strong coupling constants at the characteristic scales of the charmonium and nucleon radius, respectively.
$b=(11N_c-3N_f)/3$ is the leading coefficient of the $\beta$-function.
The quantities $T^\mu_\mu$ and $T^G_{00}$ represent the trace of the QCD energy–momentum tensor and the gluon energy density inside the proton. We further assume that the gluon energy density $T^G_{00}$ is proportional to the total proton energy density $T_{00}$, which is evaluated within the chiral quark–soliton mode($\chi$QSM)~\cite{Goeke:2007fp,Eides:2015dtr}. The parameter $\alpha_{\psi}$ denotes the chromoelectric polarizability of the charmonium state. Perturbative calculations yield $\alpha_{J/\psi}=0.2\rm GeV^{-3}$ and $\alpha_{\psi(2S)}=12\rm GeV^{-3}$~\cite{Eides:2015dtr}. In addition, the chromoelectric polarizability of the $\alpha_{J/\psi}$ has been extracted from lattice QCD in the heavy-quark limit, giving $\alpha_{J/\psi}=1.6\pm 0.8\rm GeV^{-3}$~\cite{Polyakov:2018aey}. These determinations do not yet resolve different spin configurations of the charmonium–nucleon system. More recently, the $J/\psi$–nucleon interaction has been calculated directly in HAL QCD and parametrized by a three-range Gaussian potential~\cite{Lyu:2024ttm},
\begin{eqnarray}
V(r)=\sum_i^3 c_i \exp\left(-{r^2\over b_i^2}\right).
\label{eq.potHAL}
\end{eqnarray}
The fitted parameters are $c_1=-51(-101)\rm MeV$, $c_2=-13(-33)\rm MeV$, $c_3=-22(-23)\rm MeV$, $b_1=0.09(0.13)\rm fm$, $b_2=0.49(0.44)\rm fm$, and $b_3=0.82(0.83) \rm fm$ for $J/\psi-N$ in the $S=3/2$($1/2$) state. The resulting potential, shown in Fig.~\ref{fig.pot}, is short-range and attractive.

The scattering wave function $\psi_k({\bm r})$ are calculated by solving the Schr\"odinger equation with the aforementioned potential. The radial Schr\"odinger equation for the scattering states is,
\begin{eqnarray}
\frac{d^2 u_{k,l}(r)}{dr^2} = \left( 2\mu V(r) + \frac{l(l+1)}{r^2} - k^2 \right) u_{k,l}(r),
\end{eqnarray}
where $\mu = (m_{J/\psi}m_p) / (m_{J/\psi}+m_p)$ is the reduced mass, and $u_{k,l}(r)$ represents the radial wave function in  $l-$wave scattering. For low-energy scattering, the $S-$wave channel dominates. However, as the energy increases, contributions from larger angular momenta become significant. 
Consequently, the total scattering wavefunction can be expressed as follows, $\psi_k({\bf r})=\sum_{l=0}^{l_{\rm max}}(2l+1)i^l(u_{k,l}/ r)P_{l}(\cos \theta)$, where $P_l$ denotes the Legendre polynomials. 
For low-energy scattering, the series converges relatively rapidly. In this study, we set $l_{\rm max} = 3$, and utilized CATS to solve the Schr\"odinger equation~\cite{Mihaylov:2018rva,Fabbietti:2020bfg}. 

Owing to the large mass of heavy quarks, their production predominantly occurs at the early stage of high-energy collisions through hard partonic scatterings. The formation of quarkonia involves both perturbative and nonperturbative aspects of QCD and has been studied using a variety of theoretical approaches, including the Color Evaporation Model, the Color Singlet Model, the Color Octet Model, and nonrelativistic QCD (NRQCD). More recently, we have developed a novel approach based on the quantum density matrix formalism, which provides a unified and accurate description of quarkonium production in both pp ~\cite{Song:2017phm,Zhao:2025cnp} and heavy-ion collisions~\cite{Villar:2022sbv,Song:2023zma}. This density matrix projection method has been implemented in the EPOS4 framework~\cite{Werner:2023zvo,Werner:2023fne,Werner:2023mod,Werner:2023jps}.

\begin{figure}[!htb]
    \centering
    \includegraphics[width=0.46\textwidth]{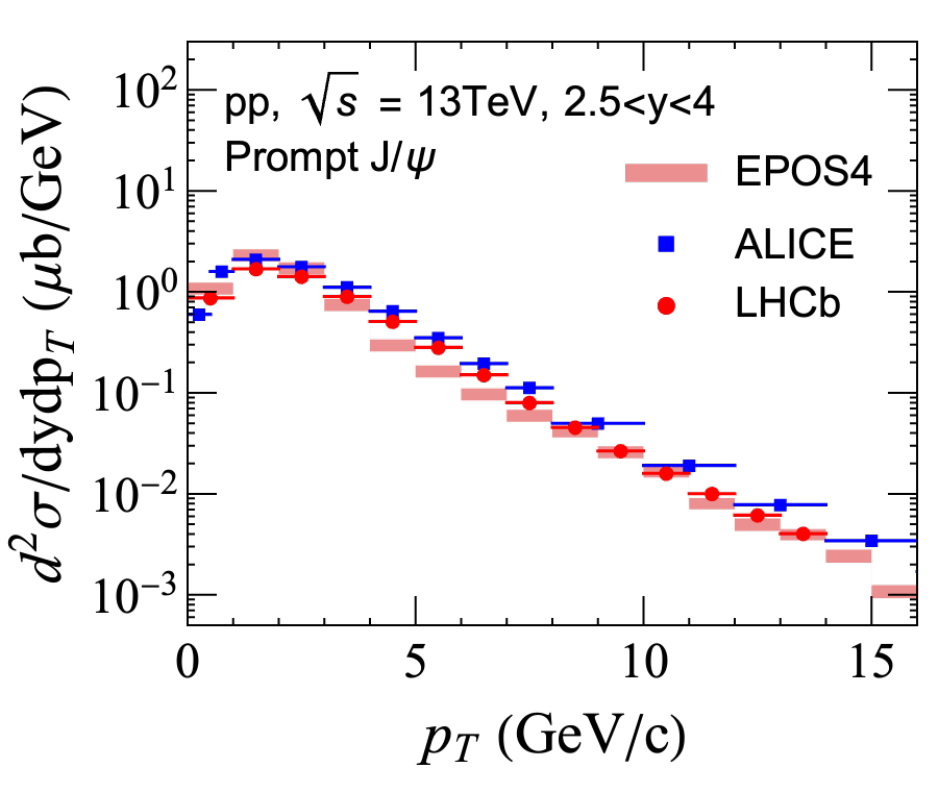}
       \includegraphics[width=0.46\textwidth]{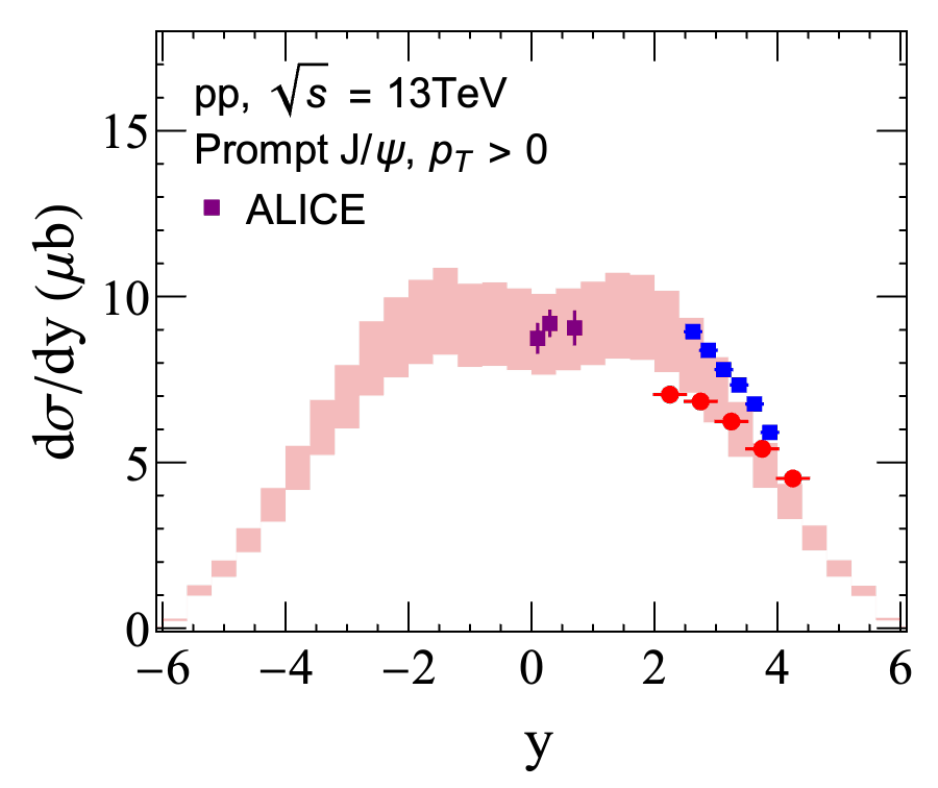}
    \caption{Transverse-momentum distribution at forward rapidity (upper panel) and rapidity distribution (lower panel) of prompt $J/\psi$ in pp collision at $\sqrt{s}=13~\rm TeV$. The experimental data is from ALICE~\cite{ALICE:2017leg,ALICE:2021dtt} and LHCb~\cite{LHCb:2015foc}. The uncertainty band arises from the variation of the width $\sigma_{c\bar c}$.}
    \label{fig.spect}
\end{figure}

Within the quantum density matrix formalism, the probability for producing a charmonium state $i$ is given by $P_i= {\rm Tr} (\rho_i \rho^{(N)})$, where $\rho_i$ denotes the density matrix of the charmonium state $i$ and $\rho^{(N)}$ represents the density matrix of the $N$ charm–anticharm pairs produced in a pp collision. Performing a partial Fourier transformation of the density matrices yields,
\begin{eqnarray}
\frac{dP_i}{{d^3\bf R}{d^3\bf P}}&=&\sum \int {d^3rd^3p \over (2\pi)^6}W_{i}^\Phi({\bf r},{\bf p})\\
&\times &\prod_{j>2} \int {d^3r_jd^3p_j \over (2\pi)^{3(N-2)}}W^{(N)}({\bf r}_1,{\bf p}_1,...,{\bf r}_N,{\bf p}_N),\nonumber
\label{eq.projection}
\end{eqnarray}
where $W_i^\Phi$ is the charmonium Wigner density and $W^{(N)}$ is the quantal density matrix in Wigner representation of the ensemble of $N$ charm-anticharm quarks. The quantal Wigner density $W^{(N)}$ is not calculable.  In numerical calculations  $W_i^\Phi W^{(N)}$ is therefore replaced by the classical phase space distribution $\langle W_i^\Phi W^{(N)}_{\rm {classical}}\rangle$ averaged over many pp events. ${\bf r}$ $({\bf R})$ and ${\bf p}$ $({\bf P})$ are the relative (center of mass) coordinate and momentum of the charm quark and anticharm pairs.
The classical phase space distributions of the heavy quarks is provided by EPOS4~\cite{Werner:2023zvo,Werner:2023fne}. We take directly the momenta given by EPOS4.
For a heavy quark pair, created at the same vertex, we assume that the relative distance  between $c$ and $\bar c$ in their center-of-mass frame, $r_{cm}$, is given by a Gaussian distribution, $r_{cm}^2\exp (-r_{cm}^2 /( 2\sigma_{\rm c\bar c})^2 )$. The relative separation is characterized by an effective width $\sigma_{c\bar c}=0.3-0.35~\rm fm$, which is the only model parameter in charmonium production. 
Instead of constructing the charmonium Wigner densities directly from their wave functions, we assume that the interaction between the $c\bar c$ pair can be approximated by a three-dimensional isotropic harmonic oscillator potential. Then, the Wigner density has an analytic form~\cite{Zhao:2025cnp},
\begin{eqnarray}
W_{\rm J/\psi}({\bf r,p})&=&8e^{-\xi} ,\\
W_{\rm \chi_c}({\bf r,p})&=& {8\over 3}e^{-\xi }(2\xi -3 ),\nonumber\\
W_{\rm \psi'}({\bf r,p})&=& {8\over 3}e^{-\xi}(3+2\xi^2-4\xi-8[p^2r^2-({\bf p}\cdot{\bf r})^2)] ),\nonumber
\end{eqnarray}
where $\xi\equiv r^2/\sigma^2+p^2 \sigma^2$. The width $\sigma$ in the Wigner density, is related to the root-mean-square radius of the charmonium state, as shown in Ref.~\cite{Zhao:2025cnp}.

Taking into account feed-down contributions from the excited states $\chi_c$ and $\psi(2S)$, with branching ratios of 30\% and 61\%, respectively~\cite{ParticleDataGroup:2022pth}, the transverse-momentum spectrum and rapidity distribution of prompt $J/\psi$ production are shown in Fig.~\ref{fig.spect}. A good agreement with experimental data in pp collisions at  $\sqrt{s}=13$~TeV is observed. This demonstrates that EPOS4 provides a reliable description of charmonium production and kinematic distributions. In EPOS4, protons are produced through either string fragmentation or fluid decay, which successfully reproduces a wide range of experimental observables~\cite{Werner:2023zvo}. This allows us to evaluate the relative distance $r$ between the charmonium and the proton in their center-of-mass frame. Since femtoscopic correlations arise predominantly from pairs with small relative momentum, we consider only charmonium–proton pairs with relative momentum below 300~MeV. 

\begin{figure}[!htb]
    \centering
    \includegraphics[width=0.46\textwidth]{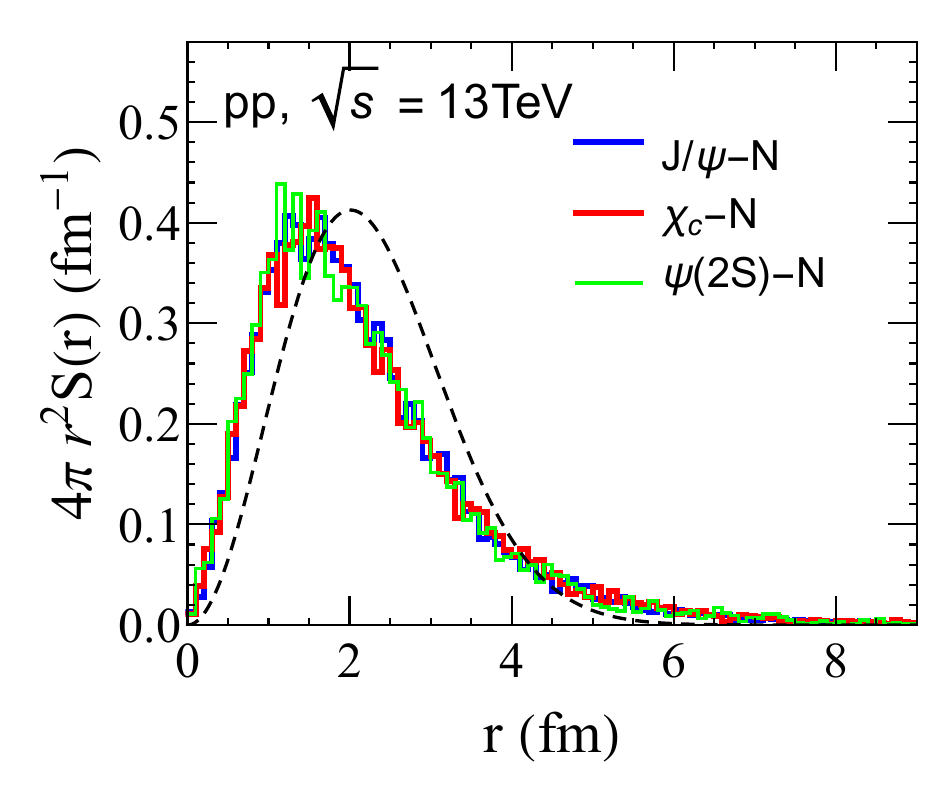}    
    \caption{The normalized emission source of charmonium-nucleon from EPOS4 in pp collisions at 13TeV. The black dashed line is the Gaussian source fit with $r_0=1.006~\rm fm$, which gives the same rms radius.}
    \label{fig.source}
\end{figure}

The normalized emission sources, $1/N dN/dr\equiv 4\pi r^2 S(r)$, are shown in Fig.~\ref{fig.source}. We find that the emission sources for $J/\psi-N$, $\chi_c-N$, and $\psi(2S)-N$ pairs are essentially identical. For comparison, we also show an isotropic Gaussian source with the same rms radius, an approximation commonly employed in femtoscopic analyses~\cite{Ohnishi:2023jlx},
$S(r) = 1/(4\pi r_{0}^{2})^{3/2}\exp(-r^{2}/(4r_{0}^{2}))$,
where $r_0$ characterizes the source size. A clear deviation of the realistic emission source from a Gaussian approximation is observed. The emission source displayed here is not a fundamental QCD observable nor a unique quantum-mechanical emission function. Rather, it represents the semiclassical freeze-out distribution of charmonium–proton pairs and serves as an input of Eq.~\eqref{eq.correlation}. 
Such a transport-based source construction has been widely employed in femtoscopic studies, where the freeze-out coordinates from event generators are directly used as the input to the Koonin–Pratt formalism~\cite{Lisa:2005dd,Li:2008qm,Wiedemann:1996ig,Lednicky:1981su,Kincses:2022eqq,Porfy:2025sev,Kincses:2025izu}. The quantitative shape of the source depends on the underlying transport model and freeze-out prescription. However, for a fixed transport model and freeze-out criterion, the source is uniquely determined by the generated event ensemble.

The correlation function, $C(k)$, is obtained by convolving the scattering wave functions with the emission source, as in Eq.~\eqref{eq.correlation}. Since the spin configuration of the $J/\psi$–proton pair cannot be resolved in current measurements, the experimentally measured correlation function corresponds to the spin-averaged value,
\begin{eqnarray}
C_{\rm ave}(k)={1\over3 }C_{S={1/2}}(k)+{2\over3}C_{S={3/2}}(k),
\end{eqnarray}
where $C_{S={1/2}}(k)$ and $C_{S={3/2}}(k)$ are correlation functions for spin 1/2 and 3/2 channels, respectively. 
\begin{figure}[!htb]
    \centering
    \includegraphics[width=0.46\textwidth]{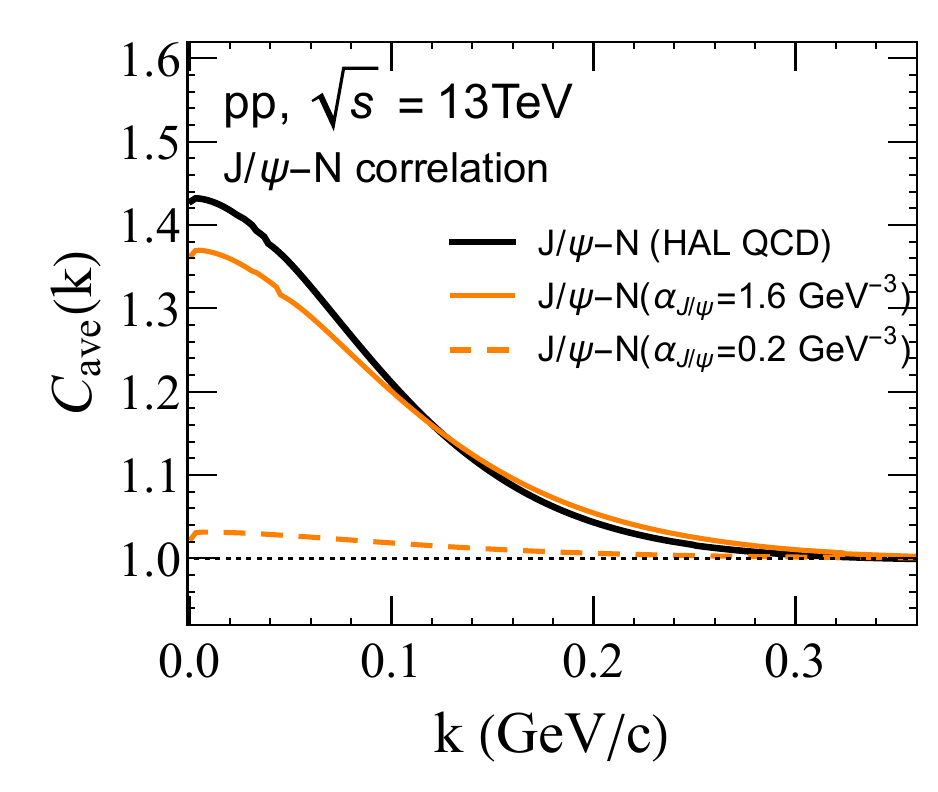}    
    \caption{Correlation function between direct $J/\psi$ and proton in pp collisions at 13 TeV. The black line is the result using the HAL QCD interaction while orange solid and dashed lines are that with interactions from the QCDME, respectively.}
    \label{fig.correlation}
\end{figure}
Fig.~\ref{fig.correlation} shows the correlation between directly produced $J/\psi$ and protons for different interactions. Since the emission source is fixed by EPOS4, the magnitude of the correlation function directly reflects the underlying interaction strength.

Experimentally, the measured (prompt) $J/\psi$ consists of approximately 60\% direct $J/\psi$, together with feed-down contributions of about 40\% from $\chi_c$ and $\psi(2S)$~\cite{Andronic:2015wma}. 
The dominant feed-down channels, $\chi_c \rightarrow J/\psi+\gamma$ and 
$\psi(2S)\rightarrow J/\psi+\pi^+\pi^-$, have significantly long lifetimes. As a result, the parent $\chi_c-N$ or $\psi(2S)-N$ pair has already left the interaction region before the decay takes place. The daughter $J/\psi$ is therefore produced at a large spatial separation from the nucleon, such that any subsequent $J/\psi-N$ interaction is strongly suppressed. Under these conditions, the feed-down contribution can be obtained by first calculating the correlation function of the parent pair and subsequently propagating it to the daughter $J/\psi-N$ system through the decay kinematics via,
\begin{eqnarray}
\label{eq.psinobs}
C_{\rm fd}^{\psi\to J/\psi}(k)&=&\int T_{\psi}(k,k_i) C_{\rm ave}^{\psi}(k_i) dk_i,\\
C_{\rm obs}(k)&=&f_{\rm dir} C^{J/\psi}(k)+\sum f_{\psi\to J/\psi} C_{\rm fd}^{\psi\to J/\psi}(k),\nonumber
\end{eqnarray}
where $T_{\psi}(k,k_i)$ denotes the decay-response matrix that maps the parent relative momentum $k_i$ of the $\psi-N$ pair to the daughter relative momentum $k$ of the $J/\psi-N$ pair. A detailed description is provided in the Appendix~\ref{ap.feeddown}. The coefficients $f_{\psi\to J/\psi}$ represent the corresponding direct and feed-down fractions of prompt $J/\psi$ production, with $f_{\rm dir}=60\%$, $f_{\chi_c\to J/\psi}=30\%$, and $f_{\psi(2S)\to J/\psi}=10\%$. 
Owing to the large threshold separations between these channels and the associated selection rules, coupled-channel effects are expected to be strongly suppressed and can be neglected.
\begin{figure}[!htb]
    \centering
    \includegraphics[width=0.46\textwidth]{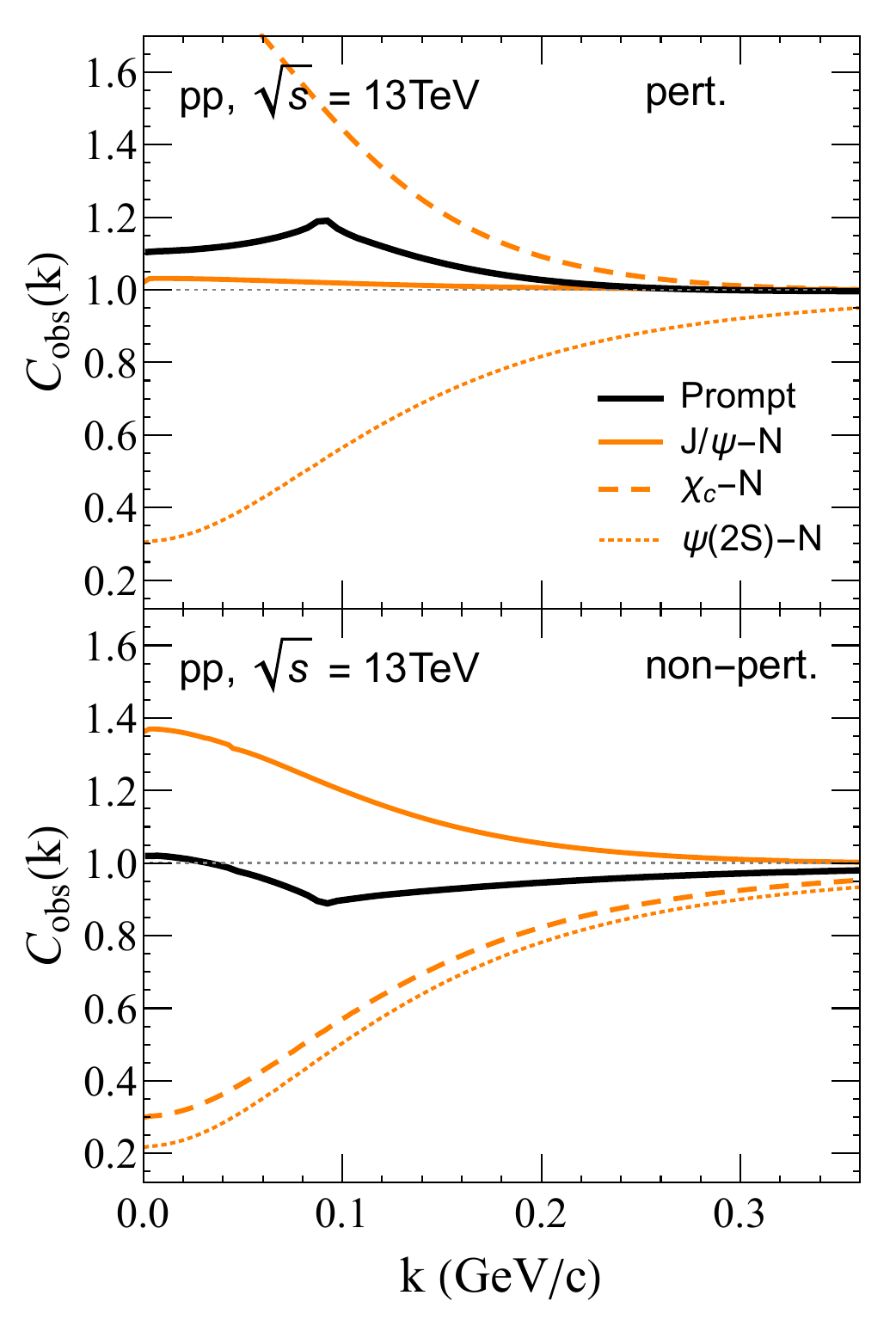}    
    \caption{Correlation functions between the direct $J/\psi$, $\chi_c$, $\psi(2S)$, prompt $J/\psi$ and the proton in pp collisions at 13 TeV.
    The interaction potentials are taken from the QCDME~\cite{Eides:2015dtr,Polyakov:2018aey}. The upper panel corresponds to perturbative chromoelectric polarizabilities, while the lower panel shows results using nonperturbative values.}
    \label{fig.correlation2}
\end{figure}

The interaction between $\chi_c-N$ and $\psi(2S)-N$ have not been simulated by HAL QCD. For applying the QCDME framework, the chromoelectric polarizability of the excited states are needed. While the chromoelectric polarizability of the $\psi(2S)$ can be extracted from its hadronic transition to the $J/\psi$, no direct determination exists for the $P$-wave charmonium states $\chi_c$. However, general considerations based on the QCD multipole expansion and pNRQCD suggest that the chromopolarizability grows with the spatial size of the quarkonium and is thus expected to be significantly larger for excited states than for the ground state, $J/\psi$.  Perturbative estimates give $\alpha_{\chi_c}=2.60~\rm GeV^{-3}$. Guided by this, we estimate the chromoelectric polarizability of an excited state $\psi$ via,
$\alpha_{\psi}=({\langle r^2 \rangle_\psi /E_{bind}^{\psi})/( \langle r^2 \rangle_{J/\psi}/E_{bind}^{J/\psi} })\alpha_{J/\psi}$,
where the rms and binding energies are obtained from the Schr\"odinger equation~\cite{Zhao:2025cnp}. Using the nonperturbative value of $\alpha_{J/\psi}$, we find $\alpha_{\chi_c}=10.37~\rm GeV^{-3}$ and $\alpha_{\psi(2S)}=47.72~\rm GeV^{-3}$. 

The correlation functions of the excited states with nucleons, as well as the prompt $J/\psi-N$ correlation function, are shown in Fig.~\ref{fig.correlation2}. Although the feed-down fractions are relatively small, the correlation of the excited states can be sizable and even smaller than unity due to their stronger interactions. Consequently, feed-down effects can noticeably modify the observed prompt $J/\psi-N$ correlation function. 
Furthermore, the prompt $J/\psi-N$ correlation exhibits a mild enhancement (or dip) around $k\simeq0.1$ GeV. This feature originates from the feed-down process $\chi_c\rightarrow J/\psi+\gamma$, which introduces a fixed recoil momentum and arises a localized kinematic structure.

In this paper, we study charmonium production and the charmonium–proton correlation function in pp collisions within the EPOS4+CATS framework. For the first time, the emission source of charmonium–proton pairs is dynamically generated in EPOS4 and exhibits a non-Gaussian structure. The use of EPOS4  provides a novel and practical femtoscopic approach to extract the charmonium–proton interaction directly from experimental correlation functions. We demonstrate that the strength of the $J/\psi-N$ interaction is clearly reflected in the corresponding correlation function. In addition, contributions from excited charmonium states, $\chi_c$ and $\psi(2S)$, interacting with the proton, are taken into account. Although their production yields are relatively small, these excited states introduce sizable uncertainties in the observed prompt $J/\psi-N$ correlation function through feed-down effects. Our results highlight the important role of excited charmonium–proton interactions and may motivate dedicated experimental and theoretical efforts in this direction. This framework also provides a unified approach to probing quarkonium–hadron interactions and simulating the corresponding correlation functions with arbitrary hadrons.

\textbf{Acknowledgment.--} 
We also acknowledge the support by the Deutsche Forschungsgemeinschaft (DFG) through the grant CRC-TR 211 "Strong-interaction matter under extreme conditions" (Project number 315477589 - TRR 211).  The computational resources utilized for this work were provided by the Center for Scientific Computing (CSC) at Goethe University Frankfurt.

\bibliographystyle{apsrev4-2}
\bibliography{refs}

\appendix
\section{Feed-down contributions}
\label{ap.feeddown}
In the present treatment, we assume that the final-state interaction acts only between the parent charmonium state and the nucleon prior to the decay. This approximation is justified by the long lifetimes of the excited charmonium states. The dominant feed-down sources, $\chi_c$ and $\psi(2S)$, decay on time scales that are significantly larger than the characteristic source size and freeze-out time of the system. As a consequence, the parent $\chi_c-N$ or $\psi(2S)-N$ pair has already left the interaction region before the decay takes place. The daughter $J/\psi$ is therefore produced at a large spatial separation from the nucleon, such that any subsequent $J/\psi-N$ interaction is strongly suppressed. Under these conditions, the observed feed-down contribution can be obtained by first calculating the correlation function of the parent pair and subsequently propagating it to the daughter $J/\psi-N$ system through the decay-response matrix. No additional $J/\psi-N$ final-state interaction after the decay is included.

In this appendix, we present a detailed treatment of the feed-down contributions using a decay-response matrix that describes the kinematic transformation from the relative momentum of the parent $\psi-N$ pair to that of the daughter $J/\psi-N$ pair, where $\psi$ denotes either $\chi_c$ or $\psi(2S)$. 
According to the Particle Data Group (PDG) compilation, the dominant decay channels are
$\chi_c \rightarrow J/\psi+\gamma$ and 
$\psi(2S)\rightarrow J/\psi+\pi^+\pi^-$. These decays modify the relative momentum of the final $J/\psi-N$ pair and therefore affect the measured femtoscopic correlation function.

For a given parent relative momentum
\begin{equation}
k_i\equiv k^*_{\psi N},
\end{equation}
the four-momenta of the parent charmonium and nucleon are initialized in the $\psi$--$N$ pair rest frame as
\begin{equation}
p_\psi^\mu=(E_\psi,0,0,k_i),
\qquad
p_N^\mu=(E_N,0,0,-k_i),
\end{equation}
with
\begin{equation}
E_\psi=\sqrt{m_\psi^2+k_i^2},
\qquad
E_N=\sqrt{m_N^2+k_i^2}.
\end{equation}

The decay $\chi_c\rightarrow J/\psi+\gamma$ is generated isotropically in the $\chi_c$ rest frame, whereas the decay $\psi(2S)\rightarrow J/\psi+\pi^+\pi^-$ is sampled according to three-body phase space. 
The daughter $J/\psi$ is then boosted back into the original $\psi$--$N$ pair frame and the relative momentum
\begin{equation}
k_f\equiv k^*_{J/\psi N}
\end{equation}
is reconstructed in the final $J/\psi$--$N$ pair rest frame.

Repeating this procedure for a large number of decay events yields the conditional probability
\begin{equation}
T_\psi(k_f,k_i)=P(k_f|k_i),
\end{equation}
which defines the decay-response matrix. Numerically,
\begin{equation}
T_\psi(k_f^j,k_i^i)=
\frac{N_\psi(k_f^j\leftarrow k_i^i)}
{\sum_j N_\psi(k_f^j\leftarrow k_i^i)},
\end{equation}
where $N_\psi(k_f^j\leftarrow k_i^i)$ denotes the number of simulated decays originating from the parent momentum bin $k_i^i$ and ending in the daughter momentum bin $k_f^j$. Here the upindex $i$ labels the parent momentum bin while $j$ labels the daughter momentum bin. 
By construction,
\begin{equation}
\sum_j T_\psi(k_f^j,k_i^i)=1.
\end{equation}
\begin{figure}[!htb]
    \centering
    \includegraphics[width=0.48\textwidth]{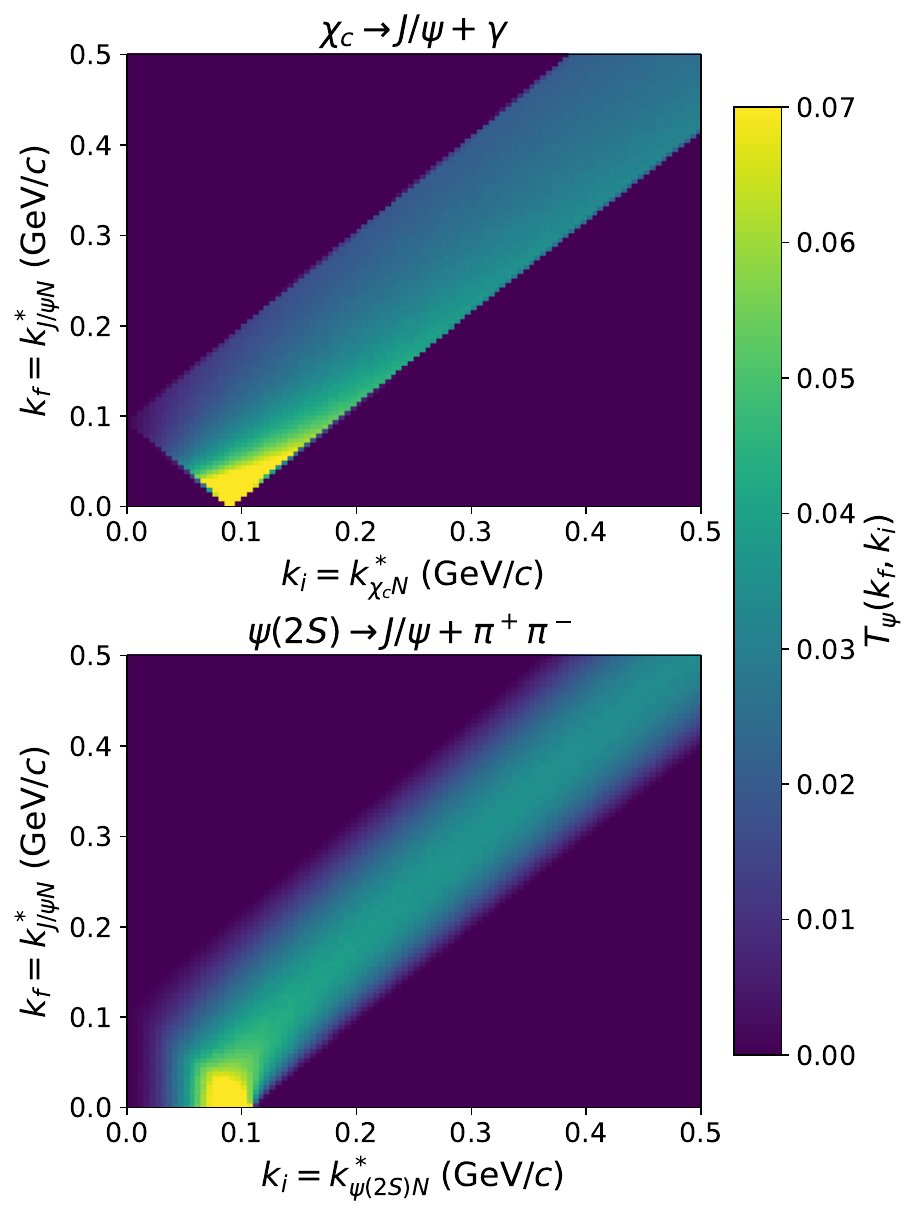}    
    \caption{Decay-response matrices $T_\psi(k_f,k_i)$ for $\chi_c\rightarrow J/\psi+\gamma$ 
(upper panel) and $\psi(2S)\rightarrow J/\psi+\pi^+\pi^-$ (lower panel).}
    \label{fig.decayT}
\end{figure}

Fig.~\ref{fig.decayT} shows the decay-response matrices $T_\psi(k_f,k_i)$ for the feed-down contributions from $\chi_c$ and $\psi(2S)$ to $J/\psi$. We can see the two-body radiative decay of the $\chi_c\to J/\psi+\gamma$ produces a sharply bounded 
kinematic region.
In the $\chi_c$ rest frame, the daughter $J/\psi$ carries a fixed recoil momentum
\begin{equation}
q_\gamma=
\frac{m_{\chi_c}^2-m_{J/\psi}^2}
{2m_{\chi_c}},
\end{equation}
which is approximately $0.39$ GeV.
The characteristic recoil scale transferred to the daughter pair is therefore
\begin{equation}
k_{\rm recoil}=
\frac{m_N}{m_N+m_{J/\psi}}q_\gamma
\simeq 0.09~{\rm GeV},
\end{equation}
This leads to a sharply bounded kinematic region. 

For three-body decays, the daughter $J/\psi$ does not carry a fixed recoil momentum, but instead samples a continuous phase space. The superposition of many recoil momenta washes out the sharp kinematic edges present in the two-body $\chi_c$ decay and produces a broad diagonal band in $T_{\psi(2S)}(k_f,k_i)$.
\begin{figure}[!htb]
    \centering
    \includegraphics[width=0.43\textwidth]{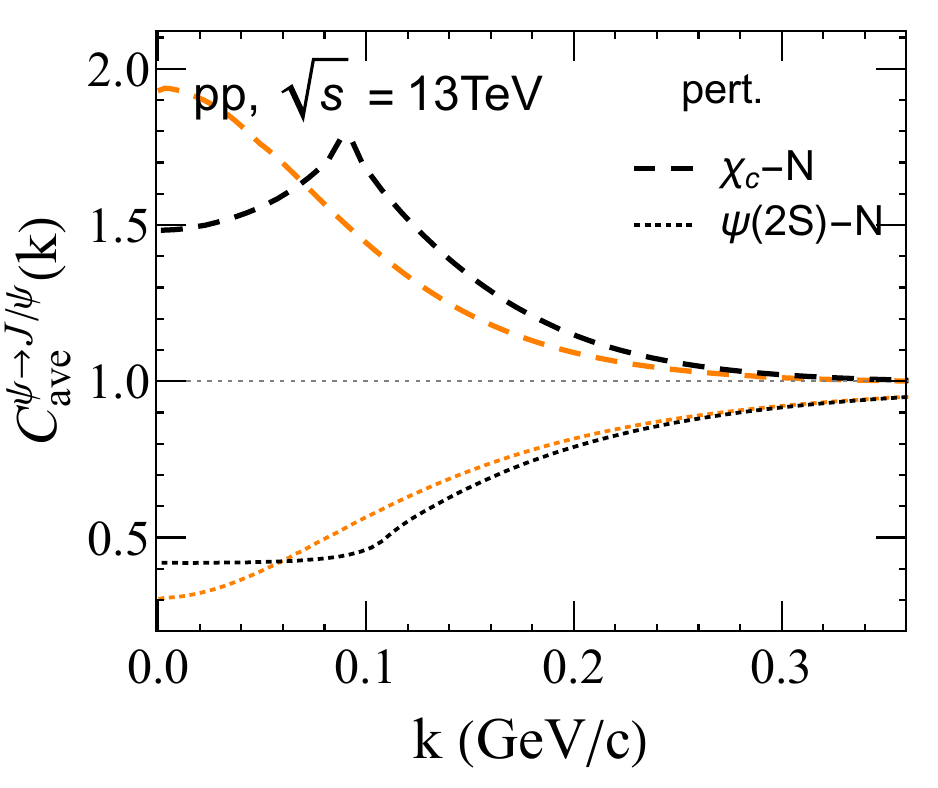}\\ \includegraphics[width=0.43\textwidth]{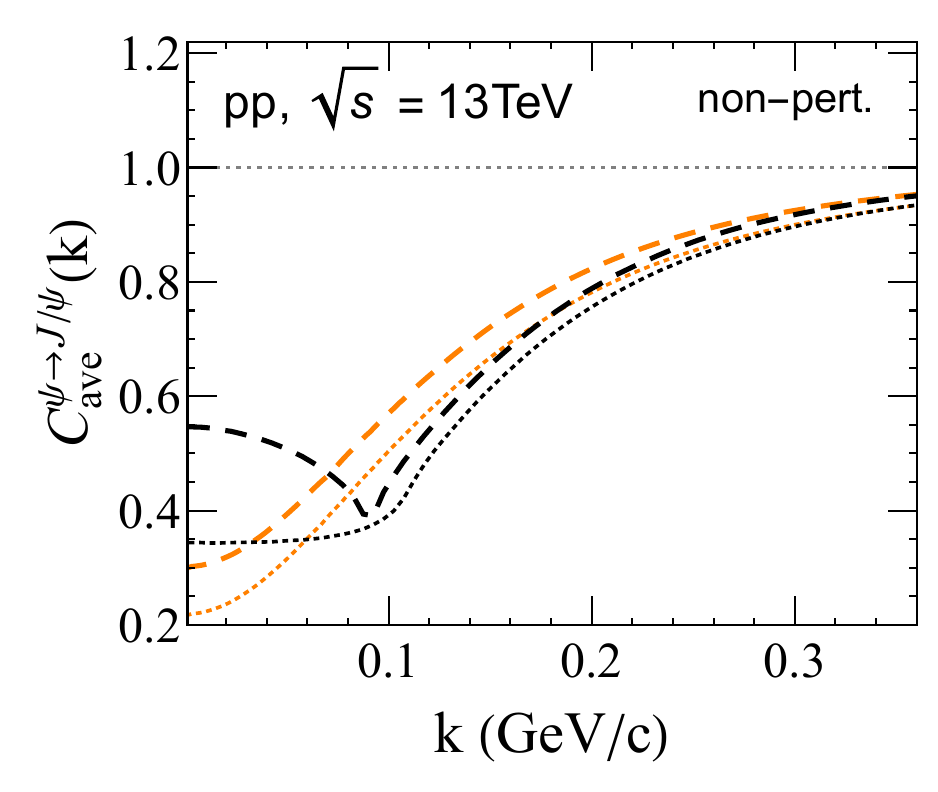}    
    \caption{Parent-state correlation functions (orange lines) with the corresponding feed-down-smeared $J/\psi-N$ correlations (black lines) obtained from the decay-response matrices. Dashed and dotted lines correspond to the $\chi_c-N$ and $\psi(2S)-N$ channels.}
    \label{fig.decayexcited}
\end{figure}

The feed-down contribution is obtained by folding the parent correlation function with the response matrix,
\begin{equation}
C_{\rm fd}^{\psi\to J/\psi}(k_f)=
\frac{
\sum_i
T_\psi(k_f,k_i)
A(k_i)
C_{\psi N}(k_i)
}
{
\sum_i
T_\psi(k_f,k_i)
A(k_i)
},
\label{eq:feeddown}
\end{equation}
where $A(k_i)$ denotes the parent pair distribution. The denominator,
\begin{equation}
R(k_f)=
\sum_i T_\psi(k_f,k_i)A(k_i),
\label{eq:response_projection}
\end{equation}
represents the projection of the response matrix onto the parent momentum distribution and characterizes the purely kinematic transformation from the parent to the daughter system.

Fig.~\ref{fig.decayexcited} compares the parent-state correlation functions (orange lines) with the corresponding feed-down-smeared $J/\psi-N$ correlations (black lines) obtained from the decay-response matrices. The left and right panels show the results for the perturbative and non-perturbative interaction scenarios, respectively. 

Overall, the decay kinematics modifies the shape of the correlation function but preserves the main interaction-induced features. The effect is most visible at low relative momentum, where the mapping between the parent and daughter systems is strongest. In particular, the feed-down contribution from the $\chi_c$ channel develops a localized structure around $k\simeq0.1$ GeV. As demonstrated by the response-matrix analysis, this feature originates from the two-body decay $\chi_c\rightarrow J/\psi+\gamma$, which introduces a characteristic recoil scale.
Consequently, the decay-response matrix exhibits a kinematic focusing around this momentum, leading to the small enhancement (or dip) visible in the smeared correlation function. This structure is physical and a consequence of the decay kinematics of the excited charmonium states and therefore should, in principle, be observable experimentally if the statistical precision is sufficient.

In contrast, the feed-down contribution from the $\psi(2S)$ channel remains considerably smoother. Since the dominant decay $\psi(2S)\rightarrow J/\psi+\pi^+\pi^-$ is a three-body process, the daughter $J/\psi$ momentum is distributed over a broad phase space rather than being fixed by a single recoil momentum. The corresponding kinematic averaging washes out localized structures and results in only a modest modification of the original correlation function. Finally, the observed prompt $J/\psi$--$N$ correlation function is constructed with Eq.~\eqref{eq.psinobs}.

\end{document}